\documentclass[reprint,nofootinbib,superscriptaddress,amsmath,amssymb,aps,prd]{revtex4-1}
\usepackage{amsmath,mathrsfs,amsbsy,color,graphicx,bm,amsthm,amsfonts}
\usepackage{bbm}
\usepackage{times}
\usepackage{units}
\usepackage{dcolumn}
\usepackage{graphicx}
\usepackage{epsfig}
\usepackage{epstopdf}
\usepackage[colorlinks,linkcolor=red,anchorcolor=green,citecolor=blue,CJKbookmarks=True]{hyperref}
\DeclareMathSymbol{\shortminus}{\mathbin}{AMSa}{"39}
\usepackage{mathrsfs}
\usepackage{braket}
\usepackage{amssymb}
\usepackage{txfonts}
\usepackage{float}
\usepackage{enumitem}
\usepackage{multirow}
\usepackage{orcidlink}

\newcommand{\pp}{\partial}

\newcommand{\meq}[1]{(\ref{#1})}

\begin{document}

\title{Entanglement degradation of static black holes in effective quantum gravity}
\author{Xiaobao Liu\orcidlink{0000-0001-6001-535X}}
\email[]{xiaobaoliu@hotmail.com} 
\affiliation{Department of Physics and Electrical Engineering, Liupanshui Normal University, Liupanshui 553004, Guizhou, China}

\author{Wentao Liu\orcidlink{0009-0008-9257-8155}}
\email[]{wentaoliu@hunnu.edu.cn (corresponding author)} 
\affiliation{Lanzhou Center for Theoretical Physics, Key Laboratory of Theoretical Physics of Gansu Province, Key Laboratory of Quantum Theory and Applications of MoE, Gansu Provincial Research Center for Basic Disciplines of Quantum Physics, Lanzhou University, Lanzhou 730000, China}

\author{Shu-Min Wu\orcidlink{0000-0001-7392-1327}}
\email[]{smwu@lnnu.edu.cn} 
\affiliation{Department of Physics, Liaoning Normal University, Dalian 116029, China}

\begin{abstract}

Quantum information science has been broadly explored in Einstein gravity and in various modified gravity theories; however, its extension to quantum gravity settings remains largely unexplored. 
Motivated by this gap, in this paper we investigate the degradation of quantum entanglement of scalar and Dirac fields in the third-type black hole geometry arising from effective quantum gravity, which incorporates generic quantum gravitational corrections beyond classical general relativity.
This quantum corrected spacetime is free of a Cauchy horizon and can be cast into a Rindler form in the near-horizon regime, allowing a direct identification of vacuum modes and a clear correspondence with the framework developed for uniformly accelerated observers. 
Within this framework, we compute the quantum entanglement and mutual information of uniformly entangled detector pairs in terms of the quantum parameter $\tilde{\zeta}$, the mode frequency $\tilde{\omega}$, and Bob’s radial position $R_0$.
The quantum parameter $\tilde{\zeta}$ consistently weakens the horizon-induced loss of correlations.
For scalar fields this effect is pronounced, producing clear departures from the classical behavior, whereas for Dirac fields the familiar correlation pattern remains intact but its degradation is noticeably reduced.
Overall, $\tilde{\zeta}$ acts as a universal protective factor against gravitational suppression of quantum correlations.
The third-type effective quantum black hole therefore provides a controlled and physically transparent arena for probing how quantum–gravity corrections influence relativistic quantum information.

\end{abstract}

\vspace*{0.5cm}

 \maketitle

\section{INTRODUCTION}

Quantum information science has reshaped our understanding of correlations, causality, and measurement across many domains of physics \cite{Boschi:1997dg,Pan:2001nrz,Hu:2018mni}.  
Over the past two decades, its concepts, particularly quantum entanglement, have become powerful probes of gravitational and relativistic phenomena \cite{Aziz:2025ypo}.  
In flat spacetime, uniformly accelerated observers experience entanglement degradation owing to the Unruh effect, a result first made precise through the analysis of mode mixing between Rindler wedges \cite{Unruh:1976db,Adesso:2007wi,Crispino:2007eb,Kollas:2022wgj}.  
When extended to curved geometries, similar techniques have revealed that event horizons play an analogous role, converting vacuum correlations into thermal noise and thereby altering the structure of bipartite quantum states \cite{Fuentes-Schuller:2004iaz,Alsing:2006cj,Fuentes:2010dt}.
These developments have led to a broad exploration of relativistic quantum information within Einstein gravity and a variety of modified gravity theories \cite{Pan:2008yr,Wang:2009qg,Martin-Martinez:2009hfq,Martin-Martinez:2010yva,Sen:2023sfb,Liu:2024wpa,Liu:2015oat,He:2016ujs,Belfiglio:2021mnr,Wu:2023sye,Wu:2023spa,Mondal:2022aev,Wu:2022lmc,Wu:2022xwy,Wu:2023spa,Wu:2023sye,Babakan:2024abb,Wu:2024qhd,Liu:2023lok,Wu:2024BF,AraujoFilho:2025rwr,AraujoFilho:2025hkm,AraujoFilho:2024ctw,Liu:2024pse,Liu:2025bzv,Qu:2024sby,Wu:2025ncd,Du:2025ipb,Wu:2025euf,Li:2025jlu,Zhang:2025jpp,Liu:2025ctf,Liu:2025lui,Feleppa:2025clx}. 
Operational tools such as Unruh-DeWitt detector models have been used to characterize how quantum fields generate correlations in relativistic settings, giving rise to research directions such as entanglement harvesting, etc \cite{Henderson:2017yuv,Cong:2018vqx,Tjoa:2020eqh,Foo:2020dzt,Zhang:2020xvo,Gallock-Yoshimura:2021yok,Zhou:2021nyv,Bueley:2022ple,Tian:2024wby,Liu:2025rks,Wu:2025qqu,Liu:2025bpp,Tang:2025mtc,Chatterjee:2025pky,Lopez-Raven:2025ehf,Teixido-Bonfill:2025wqb,Osawa:2025dja,Wang:2025lga,Huang:2025gme,Li:2025bzd,Colas:2024ysu,He:2025aht}.

However, our understanding of quantum information in genuine quantum gravity settings remains comparatively limited \cite{Lewandowski:2022zce,Yang:2022btw,Du:2025kcx,Al-Badawi:2025yqu,Panotopoulos:2025ygq,Motaharfar:2025ihv,Ai:2025myf,Li:2024ctu,Calza:2024xdh,Calza:2024fzo,Zhang:2024svj,Yang:2023gas,Yang:2025esa,Yang:2024lmj}.  
Among the many approaches to quantum gravity, effective quantum gravity models offer a particularly attractive window: they incorporate quantum corrections while retaining a covariant geometric description \cite{Zhang:2024khj,Zhang:2024ney,Zhang:2025ccx,Liu:2024soc,Sou:2024tjv,Konoplya:2024lch,Calza:2025mwn,Al-Badawi:2025rcq,Chen:2025aqh,Xamidov:2025oqx,Chen:2025ifv,Shu:2024tut,Lin:2024beb,Ban:2024qsa,Cheong:2025scf,Wang:2024iwt,Malik:2024nhy,Liu:2024wal,Malik:2024elk,Bolokhov:2024bke,Liu:2025iby,Konoplya:2025hgp,Mustafa:2025mkc,Heidari:2024bkm,Deng:2025uvp}.  
In this respect, the so-called third-type black hole solution is especially compelling \cite{Liu:2024iec,Huang:2025gia,Chen:2025baz,Paul:2025wen}.  
This geometry is free of a Cauchy horizon, exhibits a regular quantum corrected near-horizon structure, and reduces to a Rindler-type metric in the appropriate limit.  
These features make the third-type solution particularly suitable for studying how quantum gravitational corrections modify fundamental quantum processes.  
Given that black holes lie at the intersection of gravitational, quantum, and field theoretic effects, and are widely used as testbeds for probing fundamental physics, a natural question arises: how do quantum corrected black holes modify the degradation of entanglement near the event horizon?

To address this question, we compute the quantum entanglement and mutual information as explicit functions of three physical parameters: Bob’s radial distance from the horizon, the quantum parameter characterizing the underlying corrections, and the mode frequency that determines the initial entanglement between Alice and Bob.  
This analysis yields closed form expressions for the quantum correlations as functions of these parameters, allowing us to quantify precisely how quantum gravitational corrections compete with and potentially mitigate the near-horizon degradation mechanisms.

The organization of the paper is as follows.
In Sec. \ref{sec2}, we briefly introduce the effective quantum gravity model and explain why the third-type effective quantum black hole is adopted as the background for our analysis.
In Sec. \ref{sec3}, we reformulate the effective quantum black hole in Rindler coordinates and clarify the relationships among the relevant vacuum states.  
In Sec. \ref{sec4}, we derive the analytical expressions for the logarithmic negativity (measuring quantum entanglement) and mutual information of scalar and Dirac fields in this geometry, and present their behaviors as functions of the observer's radial position relative to the horizon.
Finally, Sec. \ref{sec5} summarizes our conclusions and outlines possible directions for future work.

\section{Effective quantum Black holes}\label{sec2}

Within the framework of EQG based on the Hamiltonian constraint approach, the central issue lies in preserving four-dimensional general covariance while incorporating quantum corrections to the spherically symmetric sector of vacuum gravity \cite{Zhang:2024khj,Zhang:2024ney,Zhang:2025ccx}. 
In this canonical formulation, the diffeomorphism constraint is kept in its classical form, ensuring spatial covariance, whereas the effective Hamiltonian constraint $H_{\mathrm{eff}}$ is modified to encode quantum effects through freely chosen phase space functions.  
The requirement that the algebra of constraints remains first-class imposes strong consistency conditions between $H_{\mathrm{eff}}$, the Dirac observable representing the black hole mass, and the structure function that determines hypersurface deformations.  
Solving these covariance equations yields a discrete family of covariant effective models, each characterized by a distinct realization of $H_{\mathrm{eff}}$ and a corresponding quantum parameter.
The first two types \cite{Zhang:2024khj} yield effective quantum black holes that smoothly recover the Schwarzschild solution in the classical limit, yet differ in the way the quantum parameter modifies the lapse and radial functions, leading to distinct near-horizon geometries and quasinormal spectra \cite{Liu:2024soc,Konoplya:2024lch}.
A third family \cite{Zhang:2024ney}, obtained from an alternative realization of the same covariance equations, gives rise to a new effective quantum black hole model in which the classical singularity is replaced by a regular core.
In this case, the interior region asymptotically approaches a Schwarzschild–de Sitter geometry with negative effective mass, and no Cauchy horizon appears.
The resulting causal structure differs fundamentally from the earlier two types: the spacetime contains only a single event horizon and a nonsingular interior region that smoothly transitions into an asymptotically de Sitter domain, thus avoiding both curvature singularities and inner horizon instabilities.

The metric of the effective quantum black holes by C. Zhang et al. is given by the following line element,
\begin{equation}
ds^2=-f(r)dt^2+\frac{1}{\mu(r)f(r)}dr^2+r^2d\theta^2+r^2\sin^2\theta d\varphi^2,
\end{equation}
where $f(r)$ denotes the metric function that determines the position of the event horizon, while $\mu(r)$ represents the modification factor originating from quantum gravity effects; their explicit forms differ among the three covariant types derived from the effective Hamiltonian-constraint framework.
For the first and second types \cite{Zhang:2024khj}, the effective metrics read
\begin{align}
&f_1(r)=1-\frac{2M}{r}+\frac{\zeta^2}{r^2}\left( 1-\frac{2M}{r}  \right)^2,\quad \mu_1=1,\\
&f_2(r)=1-\frac{2M}{r}, \quad~ \mu_2(r)=1+\frac{\zeta^2}{r^2}\left(1-\frac{2M}{r}\right),
\end{align}
respectively. 
Here $M$ is the ADM mass, and $\zeta$, proportional to the Planck length $\sqrt{\hbar}$ , is a quantum parameter.
The third family \cite{Zhang:2024ney}, derived from an alternative realization of the covariance equations, takes a more intricate form,
\begin{equation}\label{ds3}
\begin{aligned}
&f_3(r)=1-(-1)^n\frac{r^2}{\zeta^2}\arcsin{\left(\frac{2M\zeta^2}{r^3}\right)}-\frac{n\pi r^2}{\zeta^2}, \\
&~~\mu_3(r)=1-\frac{4M^2\zeta^4}{r^6},
\end{aligned}
\end{equation}
where $M$ and $\zeta$, the same as in the previous two types, are the mass and the quantum parameter, respectively, and $n\in \mathbb{Z} $ is an arbitrary integer, which is also regarded as a quantum parameter.
For the first two types of effective quantum black holes, the spacetime described by the metric exhibits a double-horizon structure. 
The event and inner horizons are given respectively by $r_h=2M$ and $r_m= \zeta^2/\beta-\beta/3$, with $ \beta^3=3\zeta^2\left(\sqrt{81M^2+3\zeta^2}-9M \right) $.
Considering that the main focus of this work is on the quantum information resources near the black hole horizon,  the implementations of methods \cite{Martin-Martinez:2010yva,Sen:2023sfb,Liu:2024wpa} are all based on the Damour-Ruffini approach \cite{Damour:1976jd}, which is applicable only to spacetimes with a single horizon. 
Therefore, we shall mainly concentrate on the third family of effective quantum black holes, which will hereafter be referred to simply as the effective quantum black hole.

\begin{figure}[h]
\centering 
\includegraphics[width=0.7\linewidth]{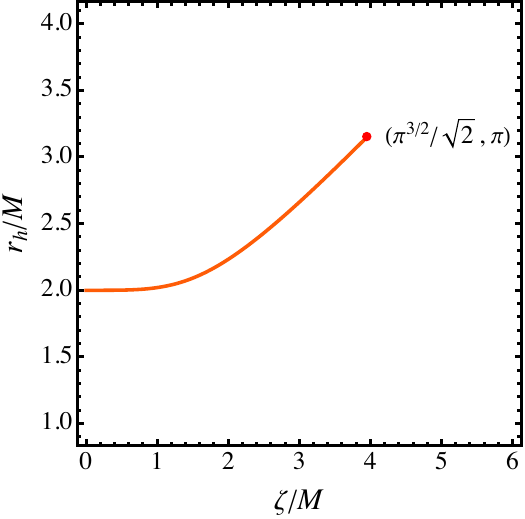} 
\caption{The relationship between the quantum parameter and the black hole event horizon is analyzed in the effective quantum black holes, with $n = 0$.}
\label{fig1}
\end{figure}
Although the horizon radius of the effective quantum black hole cannot be expressed in a closed analytic form, its behavior can be obtained numerically by solving $f_3(r)=0$ \cite{Liu:2024iec}.  
Fig. \ref{fig1} illustrates the parameter domain in which a compact object described by the metric \meq{ds3} with $n=0$ can possess an event horizon. 
The orange line segment indicates the parameter range corresponding to effective quantum black holes,  while the red dot marks the critical value at which the compact object transitions into a black hole. 
When the quantum parameter of the compact object exceeds $(\pi^{3/2}/\sqrt{2})M$, no event horizon exists. 
Conversely, for a given quantum parameter in the effective quantum black hole spacetimes, if the mass of the object is larger than $(\sqrt{2}/\pi^{3/2})\zeta$, the object forms an event horizon and can be classified as a black hole. 
For $n\neq0$, when the quantum parameter $\zeta$ is sufficiently large, an external region appears and the asymptotic behavior of the geometry approaches that of de-Sitter space. 
On the other hand, for small $\zeta$, the conventional cosmological horizon may serve as the effective event horizon of the black hole, 
in which case the entire configuration resembles a negative mass compact object.

In this work, we focus on a scenario in which two initially entangled observers, Alice and Bob, are prepared in the asymptotically flat region \cite{Fuentes-Schuller:2004iaz}. 
Alice then freely falls into the effective quantum black hole, and this physical scenario corresponds to the $n = 0$ spacetime configuration.
We therefore restate the line element as
\begin{equation}\label{dsn0}
\begin{aligned}
&ds^2=-F dt^2+\frac{\mu_3^{-1}}{F}dr^2+r^2d\theta^2+r^2\sin^2\theta d\varphi^2,\\
&F(r)=1-\frac{r^2}{\zeta^2}\arcsin\left( \frac{2M\zeta^2}{r^3} \right).
\end{aligned}
\end{equation}
For this, as shown in Fig. \ref{fig1}, we perform a numerical fitting of the relation between the horizon position and the quantum parameter, 
\begin{equation}
\begin{aligned}\label{eqrh}
\! r_h^{(5)}=&2M\!+0.033106\zeta\!-0.122211\zeta^2\!/\!M+0.139437\zeta^3\!/\!M^2\\
&-0.034416\zeta^4/M^3+0.002817\zeta^5/M^4,
\end{aligned}
\end{equation}
which provides a fifth-order polynomial fit for the horizon radius.
\begin{figure}[h]
\centering 
\includegraphics[width=0.78\linewidth]{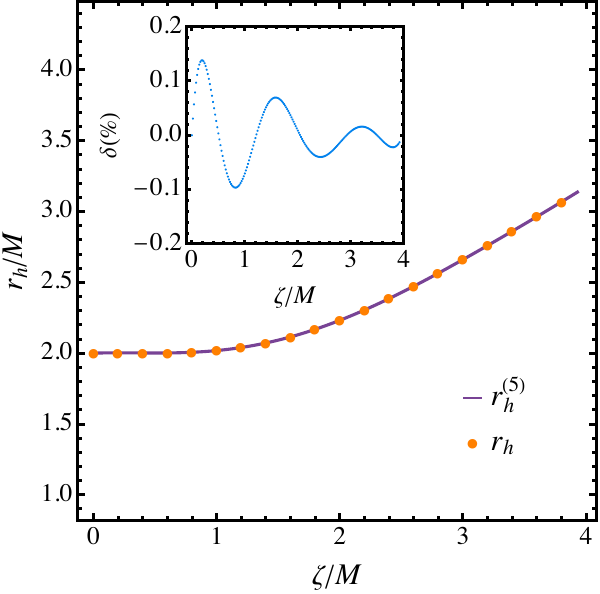} 
\caption{The comparison between the fitted horizon expression $r_h^{(5)}$ and the purely numerical result $r_h$, with the inset displaying the percentage difference between them.}
\label{fig2}
\end{figure}
To verify the reliability of our fitting results, we compare the fitted horizon expression $r_h^{(5)}$  with the purely numerical result $r_h$, as shown in Fig. \ref{fig2}. 
It can be clearly seen that the accuracy remains within $0.15\%$ throughout the valid parameter range. 
Here, $\delta(\%)$ denotes the percentage error, which is defined as  $\delta(\%) = 100(r_h^{(5)} - r_h)/r_h$.
In the following calculations, we adopt the result of equation \meq{eqrh}  as the expression for the horizon radius of the effective quantum black hole.

Now we consider the surface gravity of the effective quantum black hole, 
which can be interpreted as the proper acceleration experienced at the event horizon. 
For a static observer, the four-velocity is given by
\begin{align}
u^\mu=\left\{u^0,0,0,0\right\},
\end{align}
where $u^0$ is determined by the normalization condition $u^\mu u_\mu=-1$. 
The corresponding four-acceleration reads
\begin{align}\label{amu}
a^\nu=u^\mu\nabla_\mu u^\nu=u^\mu\partial_\mu u^\nu+\Gamma^\nu_{\mu\rho}u^\mu u^\rho.
\end{align}
In the static spacetime described by Eq.~\meq{ds3}, the surface gravity can then be obtained as
\begin{equation}
\begin{aligned}
\kappa&=\frac{1}{2}\mu_3(r_h)F'(r_h)\\
&=\frac{3M}{r_h^2}\sqrt{1\!-\!\frac{4M^2\zeta^4}{r_h^6}}
-\left(1\!-\!\frac{4M^2\zeta^4}{r_h^6} \right)\frac{r_h}{\zeta^2}\arcsin{\left(\frac{2M\zeta^2}{r_h^3}\right)}.
\end{aligned}
\end{equation}
In a general theory of gravity, when quantum particle creation is taken into account, the Hawking temperature of a black hole with constant surface gravity is given by $T=\kappa/2\pi$, as discussed in Ref. \cite{Wald:1993nt}.

\section{THE ``BLACK HOLE LIMIT'': TRANSLATION Rindler-Kruskal}\label{sec3}

Having established the reliable functional relation between the horizon radius and the quantum parameter, we now proceed to investigate the quantum information behavior in the near-horizon region, following the analysis of Refs. \cite{Martin-Martinez:2010yva,Sen:2023sfb,Liu:2024wpa}.
By employing the construction developed below for the effective quantum gravity spacetime, we are able to calculate the entanglement loss between a free-falling observer and another one located at a fixed distance from the event horizon as a function of the radial separation. 
This framework allows us to investigate the behavior of quantum correlations in the presence of effective quantum black holes. 
It is worth noting that the geometric construction proposed by S. Gangopadhyay \emph{et al.} in Ref.~\cite{Sen:2023sfb}, though insightful for spherically symmetric cases, is not directly applicable here, since the effective quantum black hole represents a different and more general spherically symmetric configuration.

For the static effective quantum black hole metric \meq{dsn0}, employing the near-horizon approximation, 
the metric function can be expanded as 
\begin{equation}
\begin{aligned}
\!\!\! F(r)=&1\!-\!\frac{r_h^2}{\zeta^2\!\!}\arcsin{\left(\frac{2M\zeta^2}{r^3_h} \right)}\!+\!(r\!-\!r_h)
\bigg[\frac{6M}{r_h^2}\bigg( 1\!-\!\frac{4M^2\zeta^4}{r_h^6} \bigg)^{\!-\!1\!/\!2}\\
&-\frac{2r_h}{\zeta^2}\arcsin{\left( \frac{2M\zeta^2}{r_h^3} \right)}\bigg]
+\mathcal{O}(r-r_h)^2,
\end{aligned}
\end{equation}
where $r_h$ denotes the event-horizon radius of the black hole. 
The first two terms on the right-hand side can be identified as $F(r_h)$. 
Since $r_h$ satisfies the condition $F(r_h)\!=\!0$, these terms vanish. 
Moreover, the coefficient of $(r - r_h)$ corresponds to $F'(r_h)$. 
Hence, the metric function near the event horizon takes the same form as that of classical black holes, namely
\begin{equation}
F(r)\simeq(r-r_h)F'(r_h).
\end{equation}
In a similar manner, the coefficient term $\mu_3(r)F(r)$ appearing in the metric component $g^{rr}$ can be expanded in a Taylor series around the event horizon, leading to
\begin{equation}
\begin{aligned}
\mu_3F\!=&\mu_3(r_h)F(r_h)\!+\!(r\!-\!r_h)\bigg[\frac{6M}{r_h^2}\sqrt{1\!-\!\frac{4M^2\zeta^4}{r_h^6}\!}+\!\frac{24M^2\zeta^4}{r_h^7}\\
&
\!-\!\bigg( 1\!+\!\frac{8M^2\zeta^4}{r_h^6} \bigg)\frac{2r_h}{\zeta^2}\arcsin{\left( \frac{2M\zeta^2}{r_h^3} \right)}
\bigg]+\mathcal{O}(r-r_h)^2\\
=&\bigg[ 1\!-\!\frac{r_h^2}{\zeta^2}\arcsin{\left(\frac{2M\zeta^2}{r_h^3}\! \right)} \bigg]\bigg[ (r\!-\!r_h)\frac{24M^2\zeta^4}{r_h^7}\!+\mu_3(r_h) \bigg]\\
&+(r-r_h)\mu_3(r_h)F'(r_h)++\mathcal{O}(r-r_h)^2,
\end{aligned}
\end{equation}
where $r_h$ satisfies $F(r_h)=0$, the related terms vanish, and we finally obtain the near-horizon relation
\begin{equation}
\mu_3(r)F(r)\simeq (r-r_h)\mu_3(r_h)F'(r_h).
\end{equation}

Following the standard approach in relativistic quantum information, we adopt a 1+1 dimensional approximation of the spacetime \cite{Fuentes-Schuller:2004iaz}.
Near the horizon, the metric then reduces to
\begin{equation}
ds^2=-(r-r_h)F'(r_h)dt^2+\frac{\mu^{-1}_3(r_h)}{(r-r_h)F'(r_h)}dr^2. 
\end{equation}
Then we define a coordinate transformation as follows:
\begin{align}\label{drdz}
d r=\frac{1}{2}\xi\mu_3(r_h) F'(r_h)d\xi.
\end{align}
By substituting equation \meq{drdz} and its integral form into the radial line element, we can obtain
\begin{equation}
\begin{aligned}
ds^2=&-\left[(c_1-r_h)F'(r_h)+\frac{1}{4}\xi^2\mu_3(r_h) F'(r_h)^2\right]dt^2\\
&+\left[1+4(c_1-r_h)/\left(\xi^2\mu_3(r_h) F'(r_h) \right) \right]^{-1}d\xi^2,
\end{aligned}
\end{equation}
where $ c_1 $ is an arbitrary integration constant.
By setting $c_1 = r_h$ and identifying the surface gravity as $ \kappa=\mu_3(r_h)F'(r_h)/2$, the metric simplifies to
\begin{align}\label{ds17}
ds^2=-\left(1\!-\!\frac{4M^2\zeta^4}{r_h^6} \right)^{-1}\xi^2 \kappa^2dt^2+d\xi^2.
\end{align}

We now introduce a static observer located at a fixed radial position $r_0$, whose proper time is denoted by $\tau$. 
For such an observer, $dr_0=0$, and the line element gives
\begin{equation}
-d\tau^2=-F(r)|_{r=r_0}dt^2+F(r)^{-1}|_{r=r_0}dr_0^2=-F(r_0)dt^2,
\end{equation}
which implies
\begin{equation}
\frac{dt}{d\tau}=\frac{1}{\sqrt{F_0}},\quad\quad t=\frac{1}{\sqrt{F_0}}\tau,
\end{equation}
where $F_0 \!\equiv\! F(r_0)$.
Substituting this relation into equation \meq{ds17},  the near-horizon metric can be recast in terms of the proper time $\tau$ as
\begin{align}\label{dstau}
ds^2=-\left(1\!-\!\frac{4M^2\zeta^4}{r_h^6} \right)^{-1}\frac{\kappa^2}{F_0}\xi^2d\tau^2+d\xi^2.
\end{align}
This is of the standard Rindler form
\begin{equation}
ds^2 = -a^2\xi^2 d\tau^2 + d\xi^2,
\end{equation}
with an effective acceleration parameter
\begin{equation}
a = \left(1-\frac{4M^2\zeta^4}{r_h^6}\right)^{-1/2}\!\! \frac{\kappa}{\sqrt{F_0}}.
\end{equation}
To elucidate the physical meaning of this parameter $a$, we next compute the proper acceleration of a static observer at $r=r_0$ in the original black hole spacetime.

For a stationary observer located at an arbitrary fixed position $r$, the value of the proper acceleration is given by
\begin{align}
a=\sqrt{a_\mu a^\mu},
\end{align}
where the four-acceleration $a^\mu$ is given by equation \meq{amu}.
Evaluating the  components of $a^\mu$ and $a_{\mu}$ for the metric \meq{dsn0} yields
\begin{align}\label{admu}
a^\mu=\left\{0,\frac{1}{2}\mu_3(r)F'(r),0,0 \right\}, && a_\mu=\left\{0,\frac{F'(r)}{2F(r)},0,0 \right\},
\end{align}
and the proper acceleration of the observer is given by
\begin{align}\label{aaaa}
a(r)=\sqrt{a^\mu a^\nu g_{\mu\nu}}=\frac{\sqrt{\mu_3(r)}F'(r)}{2\sqrt{F(r)}}.
\end{align}
For an observer located close to the event horizon ($r_0 \approx r_h$), we can use the near-horizon approximations
\begin{equation}
\begin{aligned}
F'(r)&\simeq \frac{\pp}{\pp r}\left[(r-r_h)F'(r_h)\right]\!=\!F'(r_h)\!=\!2\!\left(1\!-\!\frac{4M^2\zeta^4}{r_h^6}\right)^{\!-1}\!\kappa,\\
\mu_3(r)&\simeq1-\frac{4M^2 \zeta^4}{r_h^6}.
\end{aligned}
\end{equation}
Afterwards, proper acceleration \meq{aaaa} can be rewritten as
\begin{equation}
\!\! a(r)\simeq\left(1\!-\!\frac{4M^2\zeta^4}{r^6_h}\right)^{\!-1}\!\kappa\!\sqrt{\frac{\mu_3(r)}{F(r)}}
\simeq\left(1\!-\!\frac{4M^2\zeta^4}{r_h^6}\!\right)^{\!-1\!/\!2}\!\! \frac{\kappa}{\sqrt{F(r)}},
\end{equation}
the proper acceleration $ a_0 $ for an observer at $ r=r_0 $ is given by
\begin{align}
a_0\equiv a(r_0)=a,
\end{align}
which confirms that the acceleration parameter introduced in equation \meq{dstau} indeed represents the proper acceleration of an observer at $r_0$ in the original spacetime.
Following the framework of M. Martínez \emph{et al.} \cite{Martin-Martinez:2010yva}, with the inclusion of quantum corrections, the near-horizon geometry of the effective quantum black hole can thus be identified with a Rindler spacetime.

Next, to define the physical timelike vectors, we shift our focus to the Kruskal framework. 
Introducing the generalized null Kruskal coordinates,
\begin{equation}
 \mathcal{U}=-\frac{1}{\kappa}e^{-\kappa \big[t-{\textstyle\int} \tfrac{1}{\mu_3(r)F(r)}dr\big]},\quad
 \mathcal{V}=\frac{1}{\kappa}e^{\kappa  \big[t+{\textstyle\int} \tfrac{1}{\mu_3(r)F(r)}dr\big]},
\end{equation}
and following \cite{Martin-Martinez:2010yva}, we can define the corresponding physical timelike vectors in the three regions as
\begin{equation}
\partial_{\hat{t}} \propto \partial_\mathcal{U} + \partial_\mathcal{V},
\qquad
\partial_t \propto \mathcal{U}\partial_\mathcal{U} - \mathcal{V}\partial_\mathcal{V},
\end{equation}
together with $-\partial_t$ as the third. 
These timelike vectors define three inequivalent vacuum states, Hartle-Hawking, Boulware, and anti-Boulware, which correspond respectively to the Minkowski, Rindler, and anti-Rindler vacua,  as summarized below \cite{Martin-Martinez:2010yva}:
\begin{equation}
\begin{aligned}\label{ABBt}
&\ket{0}_\text{A}\leftrightarrow\ket{0}_\text{M}\leftrightarrow\ket{0}_\text{H},\\
&\ket{0}_\text{R}\leftrightarrow\ket{0}_\text{I~}\leftrightarrow\ket{0}_\text{B},\\
&\ket{0}_{\bar{\text{R}}}\leftrightarrow\ket{0}_\text{IV}\leftrightarrow\ket{0}_{\bar{\text{B}}}.
\end{aligned}
\end{equation}
The basis transformation between the Hartle-Hawking and Boulware modes is completely analogous to that between the Minkowski and Rindler modes, characterized by the local acceleration parameter $a_0$.

To begin with, the scalar field satisfies the Klein-Gordon equation. 
At this juncture, the metric form \meq{dsn0}, describing the effective quantum black hole spacetime, is consistent with that of the Rindler spacetime. 
This correspondence allows us to apply the standard method developed for calculating the vacuum and first-excited states of a scalar field in the Rindler case \cite{Fuentes-Schuller:2004iaz} directly to our geometry, yielding
\begin{align}\label{sk0}
\ket{0}^{\omega_i}_\text{H}=\frac{1}{\cosh\sigma_{s,i}}\sum_n \tanh^n\sigma_{s,i}\ket{n}^{\omega_i}_\text{B}
\ket{n}^{\omega_i}_{\bar{\text{B}}},
\end{align}
where $ \ket{0}_\text{H}=\otimes_j\ket{0}_\text{H}^{\omega_j} $ and 
\begin{equation}
\begin{aligned}\label{tansi}
\!\! \tanh\sigma_{\!s,i}\!=\!&\exp{\left(\!- \frac{\pi\omega_i}{a_0}\right)}
\!=\! \exp\left(-\pi\omega_i\sqrt{1\!-\!\frac{4M^2\zeta^4}{r_h^6}}\frac{\sqrt{F_0}}{\kappa}\right)\\
=&\exp{} \Biggl\{  \!-
\bigg[ \frac{3M}{r_h^2}\!-\!\!\!\sqrt{1\!-\!\frac{4M^2\zeta^4}{r^6_h}\!}\frac{r_h}{\zeta^2}\arcsin\left(\!\frac{2M\zeta^2}{r_h^3}\!\right)  \bigg]^{-\!1}\\
&\times\pi \omega_i \sqrt{1-\frac{r_0^2}{\zeta^2}\arcsin\left(\frac{2M\zeta^2}{r^3_0}\right)}  \Biggr\}.
\end{aligned}
\end{equation}
Here, $r_h$ can be directly replaced by the result from equation \meq{eqrh}.
The above result includes a quantum correction and can derive from a direct analogy with the corresponding result in the Minkowski-Rindler scenario.
The one-particle Hartle–Hawking state is generated by acting the creation operator on the vacuum, and can be expressed in the Boulware basis as
\begin{equation}\label{sk1}
\ket{1}_\text{H}^{\omega_i} = \frac{1}{\cosh^2\sigma_{s,i}} \sum_{n=0}^{\infty} \tanh^n\sigma_{s,i} \sqrt{n+1} \ket{n+1}_\text{B}^{\omega_i} \ket{n}^{\omega_i}_{\bar{\text{B}}}.
\end{equation}
The same quantization procedure applies to the Dirac field, where the Fermi–Dirac statistics modify the Bogoliubov coefficients.
The corresponding vacuum and one-particle states take the forms \cite{Alsing:2006cj}:
\begin{align}
\begin{aligned}
\ket{0}^{\omega_i}_\text{H}=&\cos\sigma_{d,i}\ket{0}^{\omega_i}_\text{B}\ket{0}^{\omega_i}_{\bar{\text{B}}}+\sin\sigma_{d,i}\ket{1}^{\omega_i}_\text{B}\ket{1}^{\omega_i}_{\bar{\text{B}}},\\
&\ket{1}^{\omega_i}_\text{H}=\ket{1}^{\omega_i}_\text{B}\ket{0}^{\omega_i}_{\bar{\text{B}}},
\end{aligned}
\end{align}
where
\begin{equation}\label{tanq}
\begin{aligned}
\!\!\tan\!\sigma_{d,i}\!=\!&\exp{} \Biggl\{  \!-
\bigg[ \frac{3M}{r_h^2}\!-\!\!\!\sqrt{1\!-\!\frac{4M^2\zeta^4}{r^6_h}\!}\frac{r_h}{\zeta^2}\arcsin\left(\!\frac{2M\zeta^2}{r_h^3}\!\right)  \bigg]^{-\!1}\\
&\times\pi \omega_i \sqrt{1-\frac{r_0^2}{\zeta^2}\arcsin\left(\frac{2M\zeta^2}{r^3_0}\right)}  \Biggr\}.
\end{aligned}
\end{equation}
The above results will serve as essential components for evaluating the effects of quantum corrections on the entanglement degradation of bosonic and fermionic fields.

\section{Quantum Entanglement in the effective quantum Black Hole spacetimes}\label{sec4}
\subsection{ Physical process and entanglement measures }

\begin{figure}[h]
\centering 
\includegraphics[width=1\linewidth]{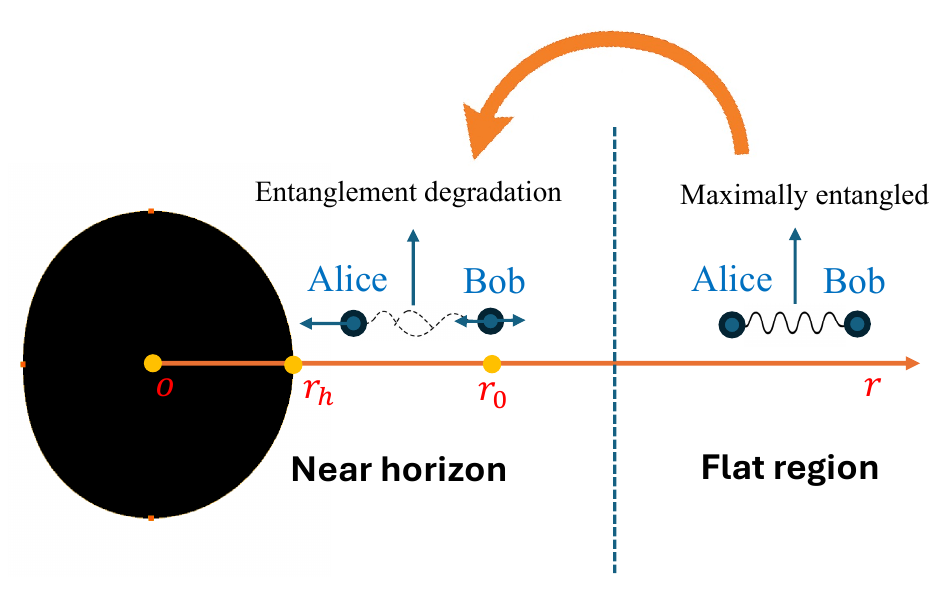} 
\caption{
Two observers (field modes), Alice and Bob, are maximally entangled in the flat region and then transported toward the near-horizon region.
As Alice freely falls into the black hole and Bob hovers outside the event horizon, their entanglement degrades due to gravitational effects.}
\label{fig3}
\end{figure}

In this subsection, we describe the physical setup used to study the degradation of quantum entanglement in the effective quantum black hole spacetime.
As illustrated in Fig. \ref{fig3}, two localized field modes (observers), conventionally labeled as Alice and Bob, are initially prepared in a maximally entangled state within the asymptotically flat region.
Subsequently, the pair is transported toward the near-horizon region, where Alice undergoes free fall across the event horizon while Bob hovers at a fixed radial position $r_0>r_h$.
The presence of the horizon causes a division of the field modes into accessible and inaccessible sectors, leading to a loss of quantum correlations when the inaccessible degrees of freedom are traced out.
This process captures the essential mechanism of entanglement degradation induced by spacetime curvature and the Hawking effect.

The initial bipartite state shared by Alice and Bob is assumed to be a maximally entangled pure state of two field modes, which takes the standard Bell form \cite{Martin-Martinez:2010yva}
\begin{align}\label{ball}
\ket{\psi}_\text{AB}=\frac{1}{\sqrt{2}}\left(\ket{0}_\text{AM}\ket{0}_\text{BM}+\ket{1}_\text{AM}\ket{1}_\text{BM}\right),
\end{align}
where Alice’s detector is sensitive only to the mode $\ket{n}\text{AM}$, and Bob’s detector is tuned exclusively to $\ket{n}\text{BM}$.
Given the correspondence previously established between the Minkowski and the black hole scenarios, the vacuum states satisfy the relation $\ket{0}\text{M}\leftrightarrow\ket{0}\text{H}$.
Using this analogy, the maximally entangled state for the observers approaching the event horizon of a black hole can be written as
\begin{align}\label{simplemax}
\ket{\psi}_\text{AB}=\frac{1}{\sqrt{2}} (\ket{0}_\text{AH} \ket{0}_\text{BH} + \ket{1}_\text{AH} \ket{1}_\text{BH}).
\end{align}
where “A” refers to Alice, who freely falls toward the event horizon, and “B” refers to Bob, who remains static at a fixed radius $r=r_0$ outside the horizon.
Both $\ket{n}\text{AH}$ and $\ket{n}\text{BH}$ denote field mode number states defined in the Hartle–Hawking basis.

To describe how the black hole geometry modifies this initially pure entangled state, we consider the tripartite system composed of Alice (A), Bob (B), and the field mode $\bar{\text{B}}$ that lies inside the event horizon and is inaccessible to external observers.
The full state of the system is described by the density operator $\rho_{\text{AB}\bar{\text{B}}}$.
For the scalar \cite{Fuentes-Schuller:2004iaz} and Dirac fields \cite{Alsing:2006cj}, the explicit forms of this tripartite density operator are given respectively by
\begin{align}
&\begin{aligned}
\!\! \rho^\text{Scalar}_{\text{AB}\bar{\text{B}}}\!=\!&\sum^\infty_{m=0}\langle m\ket{\psi_\text{s}}\langle \psi_\text{s}\ket{m}
=\frac{1}{2\cosh^2\sigma_{s,i}}\sum^\infty_{n=0}\tanh^{2n}\sigma_{s,i}\\
&\!\!\times\! \bigg[\!
\frac{\sqrt{n+1}}{\cosh\sigma_{\! s,i}} ( \ket{0~n~n}\bra{1~n\!+\!1~n}\!+\! \ket{1~n\!+\!1~n}\bra{0~n~n} )
\\&
\!\!+\! \ket{0~n~n}\bra{0~n~n} \!+\! \frac{(n+1)}{\cosh^2\! \sigma_{s,i}} \ket{1~n\!+\!1~n}\bra{1~n\!+\!1~n} \bigg],
\end{aligned}
\end{align}
and
\begin{align}
&\begin{aligned}
\!\!\rho^\text{Dirac}_{\text{AB}\bar{\text{B}}}\!=\! &\ket{\psi_\text{d}}\bra{\psi_\text{d}}
=\frac{1}{2}\Big[\cos^2\sigma_{d,i}\ket{0~0~0}\bra{0~0~0}\\
&+\sin\sigma_{d,i}\cos\sigma_{d,i}(\ket{0~0~0}\bra{0~1~1}+\ket{0~1~1}\bra{0~0~0} )\\
&+\cos\sigma_{d,i}\left(\ket{0~0~0}\bra{1~1~0}+\ket{1~1~0}\bra{0~0~0}  \right)\\
&+\sin\sigma_{d,i}\left(\ket{0~1~1}\bra{1~1~0}+\ket{1~1~0}\bra{0~1~1}  \right)\\
&+\sin^2\sigma_{d,i}\ket{0~1~1}\bra{0~1~1}+\ket{1~1~0}\bra{1~1~0}\Big].
\end{aligned}
\end{align}
Since an external observer cannot access the field modes inside the horizon, the physical states relevant for observable correlations are obtained by tracing over the inaccessible degrees of freedom.
The reduced density matrices for different bipartitions are thus given by
\begin{align}
\begin{aligned}
\rho_\text{AB}=\text{Tr}_{\bar{\text{B}}}\rho_{\text{AB}\bar{\text{B}}},&&
\rho_{\text{A}\bar{\text{B}}}=\text{Tr}_{{\text{B}}}\rho_{\text{AB}\bar{\text{B}}},&&
\rho_{\text{B}\bar{\text{B}}}=\text{Tr}_{{\text{A}}}\rho_{\text{AB}\bar{\text{B}}}.
\end{aligned}
\end{align}
These reduced states encode the correlations accessible to different observers and serve as the foundation for quantifying the degradation of entanglement induced by the presence of the black-hole horizon.
In particular, the $\text{AB}$ bipartition corresponds to an inertial observer paired with the field modes accessible to the accelerated observer, while the $\text{A}\bar{\text{B}}$ bipartition involves Alice and the field modes that Bob cannot access due to the causal structure of spacetime.
Classical communication is possible only within the $\text{AB}$ and $\text{A}\bar{\text{B}}$ bipartitions \cite{Alsing:2006cj}, which are the only configurations that permit quantum information tasks.
In the following analysis, we adopt two complementary information theoretic quantities to characterize the correlations between the subsystems: the logarithmic negativity and the mutual information.

The logarithmic negativity, denoted by $\mathcal{N}$, is an entanglement monotone that captures the amount of distillable quantum entanglement.
It is defined as \cite{Vidal:2002zz,Diaz:2023jrf,Xu:2024eqg}
\begin{align}
\mathcal{N}(\rho_{\text{AB}})=\log_2 \left|\left|  \rho_\text{AB}^{T_\text{A}}\right|\right|_1,
\end{align}
for the maximally entangled bipartite state, for both scalar and Dirac fields.
Here, $T_\text{A}$ denotes the partial transpose with respect to subsystem A, and $\|\cdot\|_1=\mathrm{Tr}\!\sqrt{\rho^\dagger\rho}$ is the trace norm.
For scalar fields, our focus is on $\rho_\text{AB}$, the density operator characterizing the bipartite system of Alice and Bob.
We will analyze the entanglement degradation as a function of Bob’s position, as described by
\begin{equation}\label{NS1}
\begin{aligned}
\mathcal{N}\left(\rho^\text{Scalar}_\text{AB}\right)=&\log_2||\rho^\text{Scalar,$\text{T}_\text{A}$}_\text{AB}||_1\\
=&\log_2\left[\frac{1}{2\cosh^2\sigma_{s,i}}+\sum^\infty_{n=0}\frac{\tanh^{2n}\sigma_{s,i} \sqrt{\mathcal{C}_n}}{2\cosh^2\sigma_{s,i}}  \right],
\end{aligned}
\end{equation}
with
\begin{equation*}
\mathcal{C}_n=\left(\tanh^2\sigma_{s,i}+\frac{n}{\sinh^2\sigma_{s,i}}\right)^2+\frac{4}{\cosh^2\sigma_{s,i}}.
\end{equation*}
For Dirac fields, due to the physical characteristics of entanglement redistribution \cite{Wang:2010bf}, we focus on both $ \rho_{\text{AB}} $ and $ \rho_{\text{A}\bar{\text{B}}} $, the density operators characterizing the bipartite systems of Alice-Bob and Alice-antiBob, respectively.
The logarithmic negativity in this case are give by:
\begin{equation}\label{ND1}
\begin{aligned}
\mathcal{N}\left(\rho^\text{Dirac}_\text{AB}\right)=&\log_2||\rho^\text{Dirac,T}_\text{AB}||_1=\log_2\left(1+\cos^2\sigma_{d,i}\right),
\end{aligned}
\end{equation}
\begin{equation}\label{ND2}
\begin{aligned}
\mathcal{N}\left(\rho^\text{Dirac}_{\text{A}\bar{\text{B}}}\right)=&\log_2||\rho^\text{Dirac,T}_{\text{A}\bar{\text{B}}}||_1=\log_2\left(1+\sin^2\sigma_{d,i}\right).
\end{aligned}
\end{equation}

On the other hand, the mutual information, which gives an idea of the total amount of correlation, is defined for a bipartite system AB as \cite{Fuentes-Schuller:2004iaz}
\begin{align}
I(\rho_\text{AB})=S(\rho_\text{A})+S(\rho_\text{B})-S(\rho_\text{AB}),
\end{align}
where $ S(\rho)=-\text{Tr}(\rho \log_2\rho) =-\sum_i\lambda_{i}\log_2\lambda_{i}$ is the entropy of the density matrix $ \rho $, with $\lambda_{i}$ being its eigenvalue.
For scalar fields, the entropy of the joint state is
\begin{equation}
\begin{aligned}
S(\rho^\text{Scalar}_\text{AB})=&-\sum_{n=0}^{\infty}\frac{\tanh^{2n}\sigma_{s,i}}{2\cosh^2\sigma_{s,i}}\left(1+\frac{n+1}{\cosh^2\sigma_{s,i}}\right)\\
&\times\log_2\left[\frac{\tanh^{2n}\sigma_{s,i}}{2\cosh^2\sigma_{s,i}}\left(1+\frac{n+1}{\cosh^2\sigma_{s,i}}\right)\right].
\end{aligned}
\end{equation}
The density matrix for Bob is obtained by tracing out Alice's states, and its entropy is
\begin{equation}
\begin{aligned}
S(\rho^\text{Scalar}_\text{B})=&-\sum_{n=0}^{\infty}\frac{\tanh^{2n}\sigma_{s,i}}{2\cosh^2\sigma_{s,i}}\left(1+\frac{n}{\sinh^2\sigma_{s,i}}\right)\\
&\times\log_2\left[\frac{\tanh^{2n}\sigma_{s,i}}{2\cosh^2\sigma_{s,i}}\left(1+\frac{n}{\sinh^2\sigma_{s,i}}\right)\right].
\end{aligned}
\end{equation}
Given that $ S(\rho^\text{Scalar}_\text{A})=1 $, the mutual information is
\begin{equation}\label{IS1}
\begin{aligned}
I(\rho_\text{AB}^\text{Scalar})=&1-\frac{1}{2}\log_2\tanh^2\sigma_{s,i}-\sum_{n=0}^{\infty}\frac{\tanh^{2n}\sigma_{s,i}}{2\cosh^2\sigma_{s,i}} \bar{\mathcal{C}}_n,
\end{aligned}
\end{equation}
with
\begin{align*}
\bar{\mathcal{C}}_n=&\left(1+\frac{n}{\sinh^2\sigma_{s,i}}\right)\log_2\left(1+\frac{n}{\sinh^2\sigma_{s,i}}\right)\\
&-\left(1+\frac{n+1}{\cosh^2\sigma_{s,i}}\right)\log_2\left(1+\frac{n+1}{\cosh^2\sigma_{s,i}}\right).
\end{align*}
For Dirac fields, the mutual information corresponding to the bipartite systems AB and  $ \text{A}\bar{\text{B}} $ can be expressed as follows:
\begin{equation}\label{ID1}
\begin{aligned}
I(\rho^\text{Dirac}_{\text{A}\text{B}})=&1-\frac{1}{2}\cos^2\sigma_{d,i}\log_2\left(\frac{\cos^2\sigma_{d,i}}{2}\right)\\
&-\left(1-\frac{1}{2}\cos^2\sigma_{d,i}\right)\log_2\left(1-\frac{1}{2}\cos^2\sigma_{d,i}\right)\\
&+\frac{1}{2}\left(1+\cos^2\sigma_{d,i}\right)\log_2\left(\frac{1+\cos^2\sigma_{d,i}}{2}\right)\\
&+\frac{1}{2}\left(1-\cos^2\sigma_{d,i}\right)\log_2\left( \frac{1-\cos^2\sigma_{d,i}}{2} \right),
\end{aligned}
\end{equation}
and
\begin{equation}\label{ID2}
\begin{aligned}
I(\rho^\text{Dirac}_{\text{A}\bar{\text{B}}})=&1-\frac{1}{2}\sin^2\sigma_{d,i}\log_2\left(\frac{\sin^2\sigma_{d,i}}{2}\right)\\
&-\left(1-\frac{1}{2}\sin^2\sigma_{d,i}\right)\log_2\left(1-\frac{1}{2}\sin^2\sigma_{d,i}\right)\\
&+\frac{1}{2}\left(1+\sin^2\sigma_{d,i}\right)\log_2\left(\frac{1+\sin^2\sigma_{d,i}}{2}\right)\\
&+\frac{1}{2}\left(1-\sin^2\sigma_{d,i}\right)\log_2\left( \frac{1-\sin^2\sigma_{d,i}}{2} \right).
\end{aligned}
\end{equation}
Using the above expressions \meq{NS1}, \meq{ND1}, \meq{ND2} \meq{IS1},\meq{ID1} and \meq{ID2}, we will present the corresponding numerical results in the next subsection.

\subsection{Results}
In this subsection, we present the numerical results for the entanglement behavior of both bosonic and fermionic fields in the effective quantum black hole spacetime.
Using the expressions derived in the previous subsection, 
we evaluate the logarithmic negativity and mutual information for the bipartite systems AB and $\text{A}\bar{\text{B}}$ as functions of the mode frequency $ \omega_i $, the initial radial position $r_0$, and the quantum corrected geometric parameters introduced by the effective metric.
These quantities allow us to systematically examine how quantum corrections to the horizon structure influence the redistribution and degradation of correlations when Alice approaches the event horizon while Bob remains static outside.

Before presenting the numerical results, it is necessary to clearly specify the physical quantities involved in the analysis, all of which are taken to be dimensionless.
Using Bob’s locally measured mode frequency and his radial position expressed in units of the black hole horizon radius $r_h$, we introduce the following dimensionless parameters:
\begin{align}
\tilde{\omega}=\omega_i M,&& \tilde{\zeta}=\zeta/M, && R_0=r_0/r_h.
\end{align}
With these definitions, the dynamics of the detector pair becomes entirely controlled by the dimensionless triplet $(\tilde{\omega},\tilde{\zeta},R_0)$.
This parametrization isolates the physical effects induced by the quantum corrected geometry from trivial mass rescalings and allows for a direct comparison across different black-hole backgrounds.
In what follows, we compute the reduced density matrices of the bipartite system and evaluate the corresponding entanglement measures, focusing on their dependence on $(\tilde{\omega},\tilde{\zeta},R_0)$.
To make the physical distinctions transparent, we discuss the scalar and Dirac fields separately.
For each field, our analysis proceeds in two steps: we first investigate the degradation of quantum entanglement, and subsequently examine the corresponding mutual information.
This leads to four subsections: scalar field entanglement, scalar field mutual information, Dirac field entanglement, and Dirac field mutual information.

\begin{figure}[h]
\centering 
\includegraphics[width=0.48\linewidth]{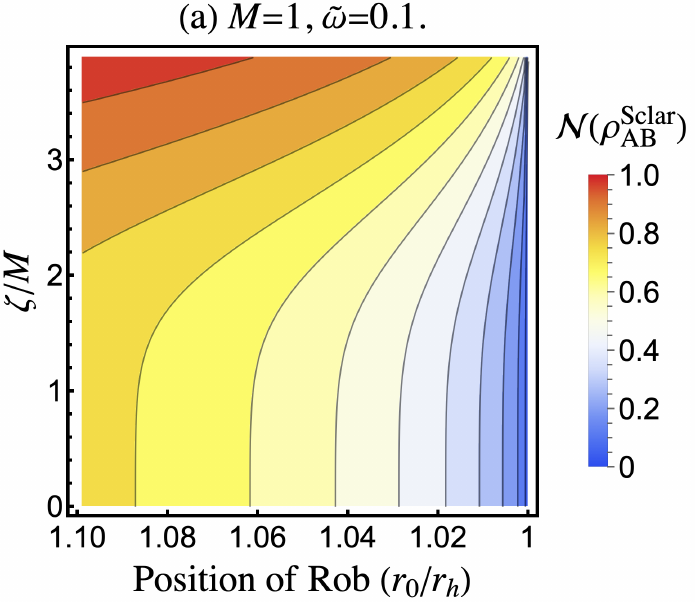} 
\includegraphics[width=0.48\linewidth]{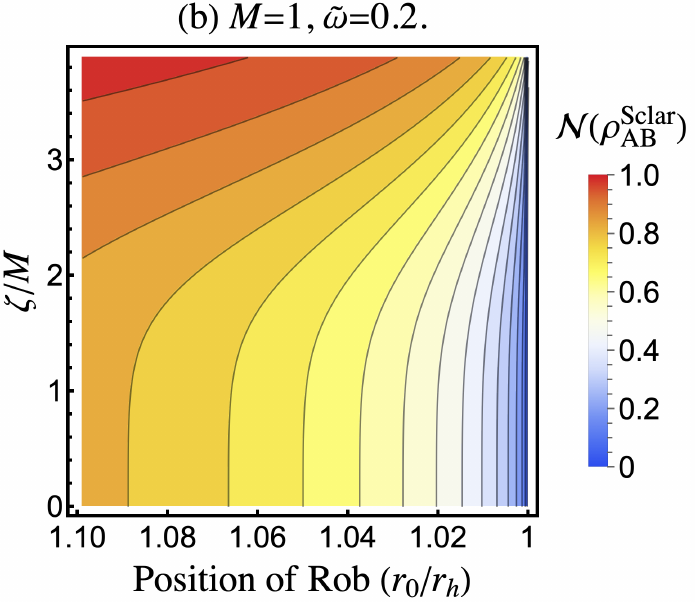}  
\includegraphics[width=0.48\linewidth]{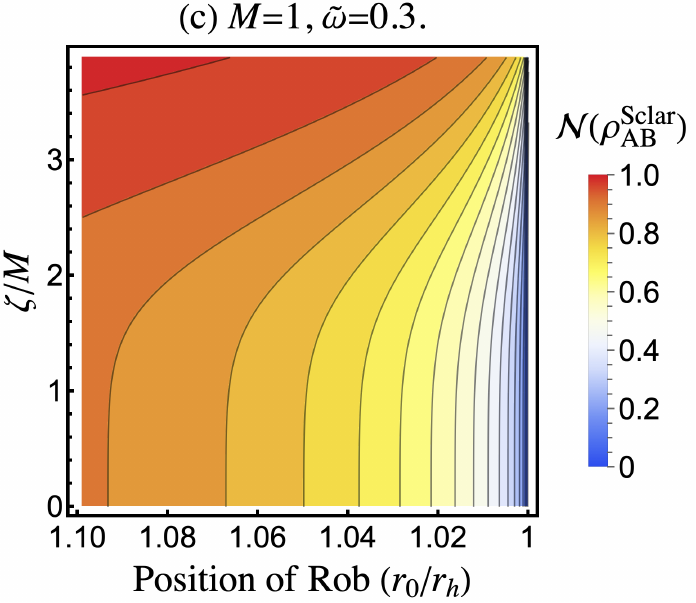}  
\includegraphics[width=0.48\linewidth]{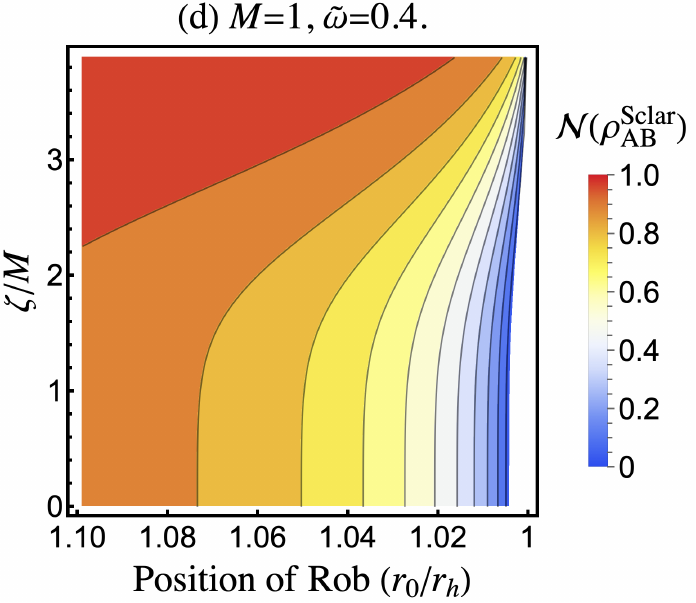}   
\caption{
Scalar field: The entanglement of the Alice-Bob system is analyzed as a function of Bob’s position $R_0 = r_0/r_h$ and the quantum parameter $\tilde{\zeta}=\zeta/M$ for different values of $\tilde{\omega}=\omega_i M$.
The entanglement vanishes as Bob approaches the horizon $r_h$, and no entanglement is created between Alice and antiBob. }
\label{fig4}
\end{figure}
\begin{figure}[h]
\centering 
\includegraphics[width=0.48\linewidth]{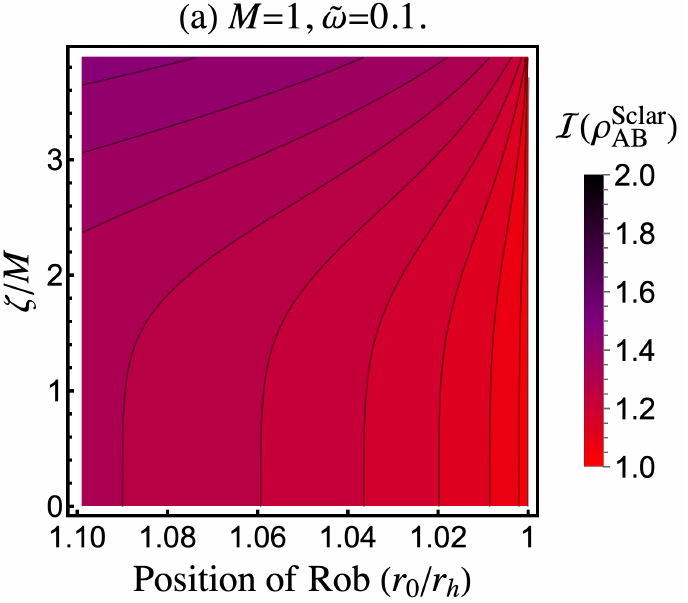} 
\includegraphics[width=0.48\linewidth]{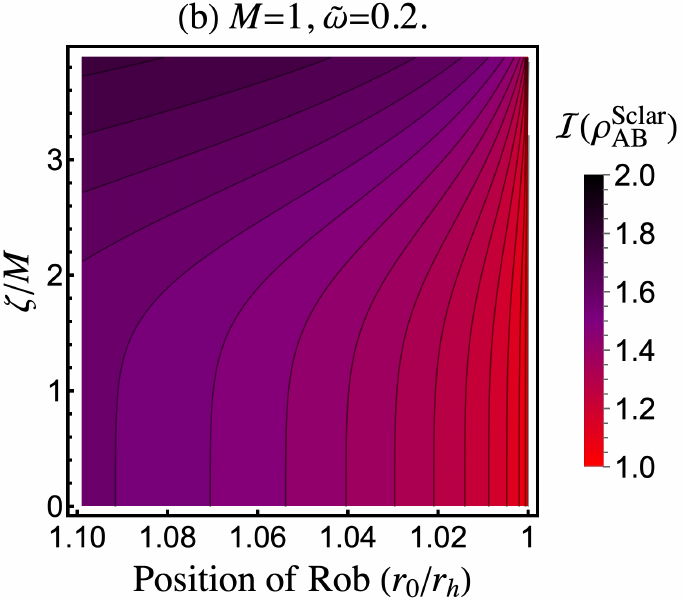}  
\includegraphics[width=0.48\linewidth]{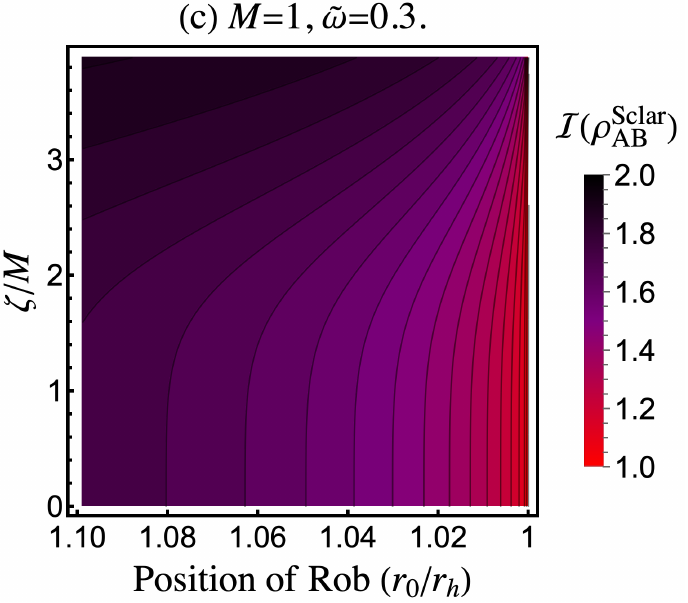}  
\includegraphics[width=0.48\linewidth]{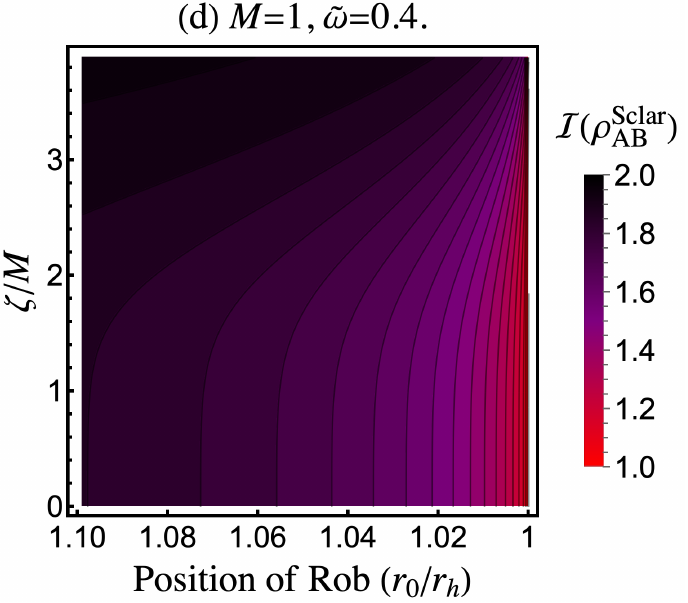}   
\caption{
Scalar field: The mutual information of the Alice–Bob system is analyzed as a function of Bob’s position $R_0 = r_0/r_h$ and the quantum parameter $\tilde{\zeta}=\zeta/M$ for different values of $\tilde{\omega}=\omega_i M$.
}
\label{fig5}
\end{figure}
Figs. \ref{fig4} and \ref{fig5} display the behavior of the scalar field entanglement and mutual information as functions of Bob’s radial position $R_0=r_0/r_h$ and the quantum parameter $\tilde{\zeta}$, for several representative detector frequencies $\tilde{\omega}$.
Both quantities exhibit clear and consistent patterns governed by the interplay between the near-horizon geometry and the quantum corrections encoded in $\tilde{\zeta}$.
For all values of $\tilde{\omega}$, the scalar-field entanglement increases monotonically with $\tilde{\zeta}$.
This indicates that the quantum corrections to the geometry mitigate the gravity-induced degradation of entanglement.
This tendency is consistent with the modified Unruh/Hawking-type thermalization in the quantum corrected background: larger $\tilde{\zeta}$ corresponds to a lower effective local temperature, which reduces thermal mode mixing and therefore slows the loss of bipartite entanglement.
At fixed $\tilde{\zeta}$, the entanglement decreases markedly as Bob approaches the horizon $(R_0\!\to\!1)$.
This reflects the enhanced near-horizon redshift, which strengthens the field fluctuations and the corresponding mode-mixing channel responsible for degrading the shared entanglement.
The contour gradients along the $R_0$ direction become steeper for larger $\tilde{\omega}$, showing that high-frequency modes respond more sensitively to the near-horizon environment.
The scalar field mutual information exhibits a very similar absolute variation pattern.
In the figs \ref{fig5}, black regions correspond to $I\simeq 2$, while red regions correspond to $I\simeq 1$.
At fixed $\tilde{\zeta}$, the mutual information decreases as Bob approaches the horizon, again indicating that near-horizon effects efficiently reduce total (quantum + classical) correlations.
For any fixed $R_0$ away from the horizon, the mutual information increases monotonically with $\tilde{\zeta}$, in close parallel with the behavior of entanglement.
This is again explained by the fact that larger $\tilde{\zeta}$ lowers the effective local temperature in the quantum corrected geometry, resulting in weaker noise and a reduced loss of correlations.
Thus, when examined through its absolute variation from $I=2$ to $I=1$, the mutual information behavior closely mirrors that of the negativity.
Finally, increasing $\tilde{\omega}$ reduces the overall magnitude of the mutual information, with the high-value region shrinking as $\tilde{\omega}$ grows.
High-frequency detectors therefore carry less total correlation even though their quantum entanglement is better protected.

\begin{figure}[h]
\centering 
\includegraphics[width=0.48\linewidth]{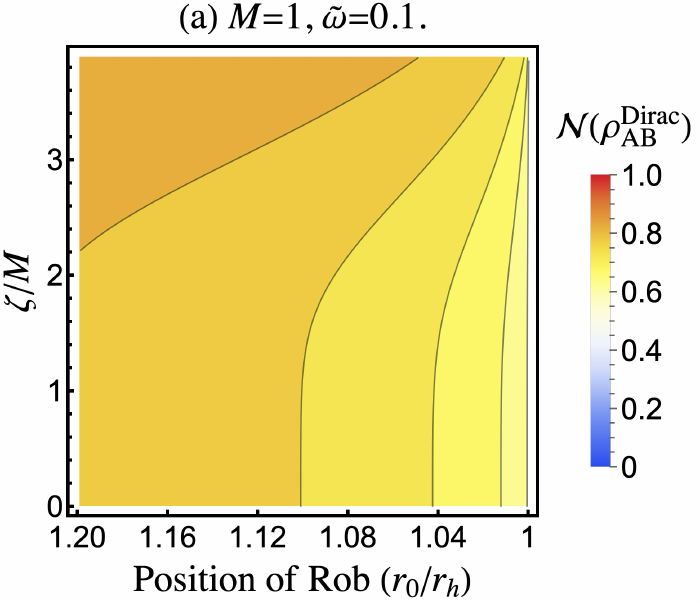} 
\includegraphics[width=0.48\linewidth]{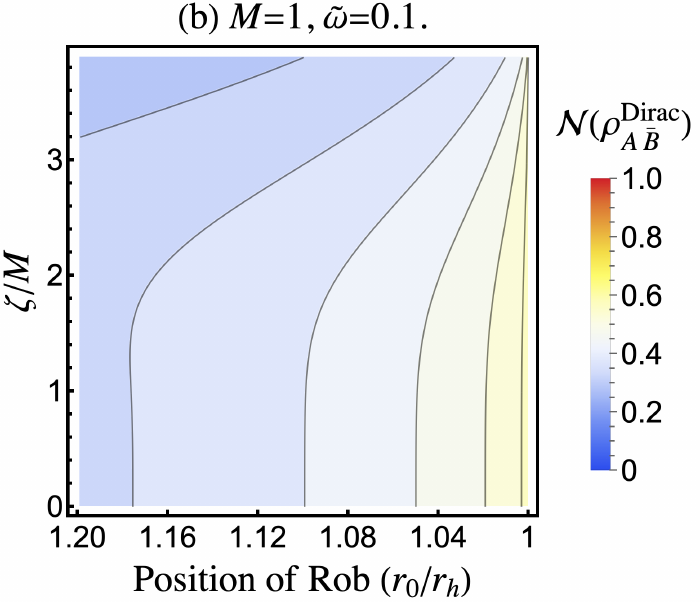}  
\includegraphics[width=0.48\linewidth]{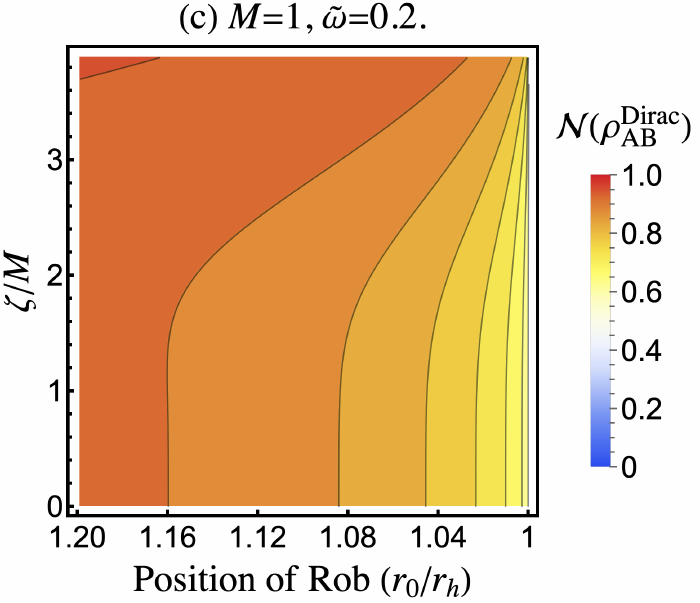}  
\includegraphics[width=0.48\linewidth]{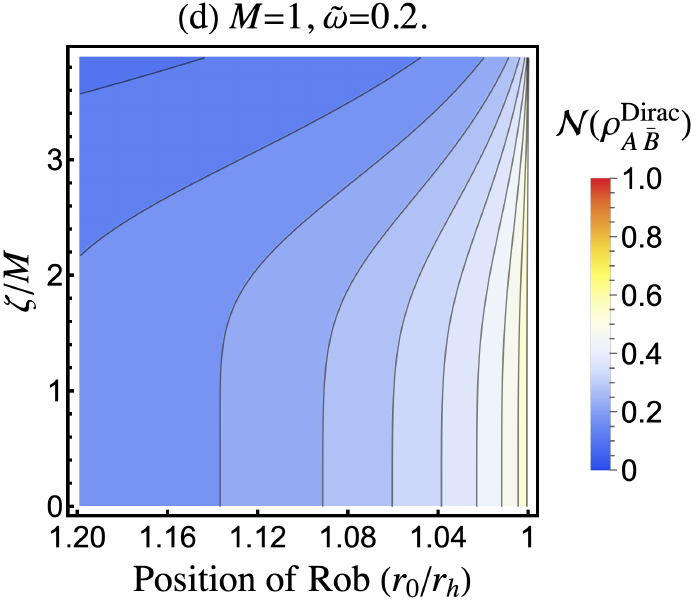}   
\caption{
Dirac field: The entanglement of the Alice-Bob system (Left) and the Alice-antiBob system (Right) as a function of Bob’s position $R_0 = r_0/r_h$ and the quantum parameter $\tilde{\zeta}=\zeta/M$ for different values of $\tilde{\omega}=\omega_i M$.
}
\label{fig6}
\end{figure}
\begin{figure}[h]
\centering 
\includegraphics[width=0.48\linewidth]{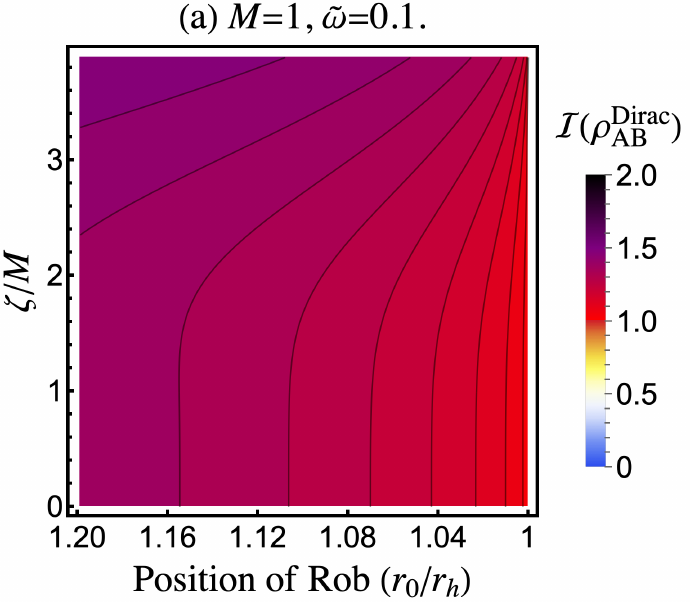} 
\includegraphics[width=0.48\linewidth]{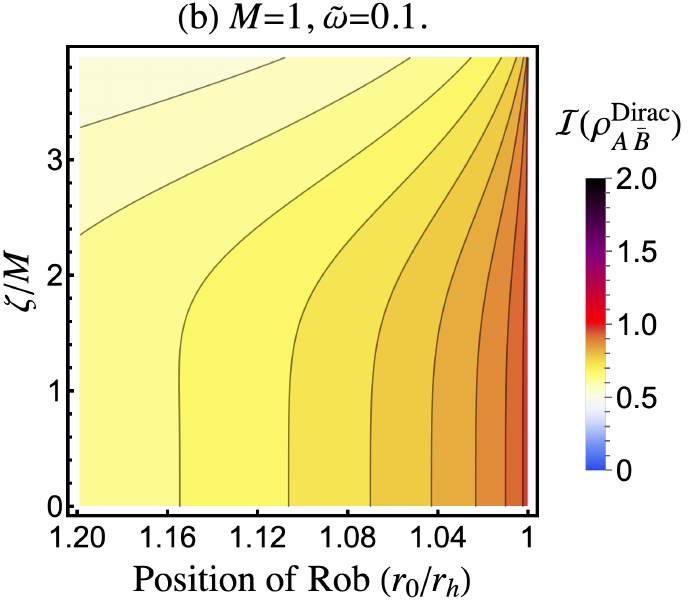}  
\includegraphics[width=0.48\linewidth]{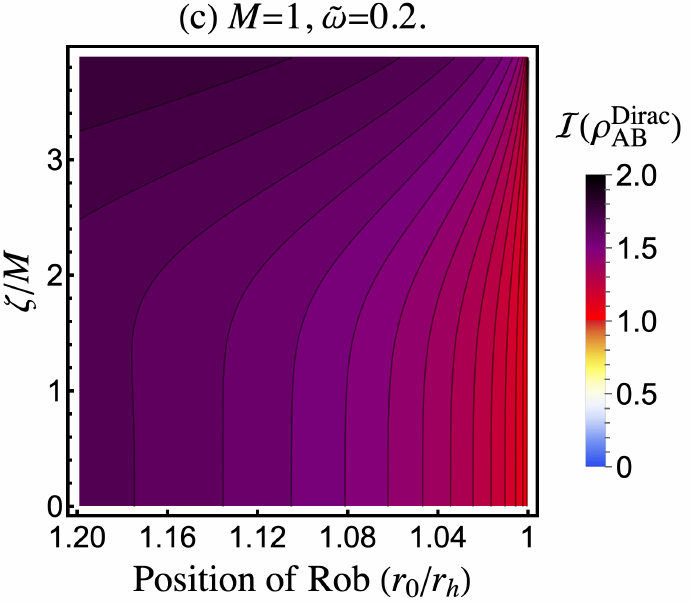}  
\includegraphics[width=0.48\linewidth]{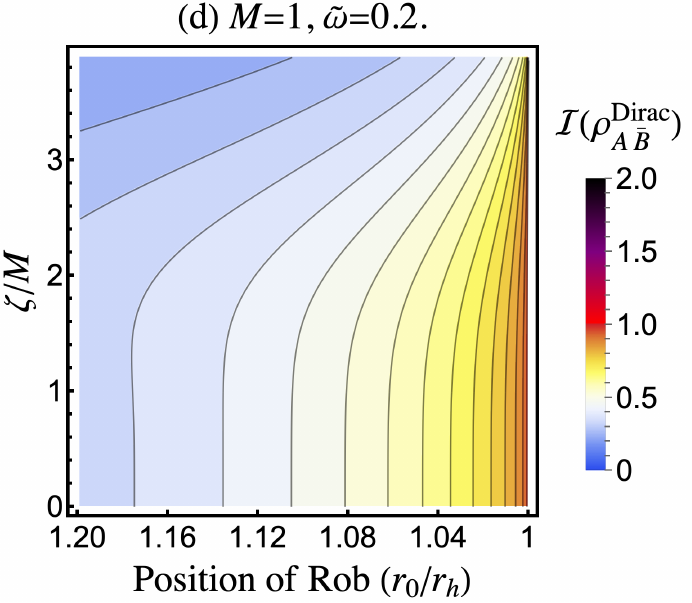}   
\caption{
Dirac field: The mutual information of the Alice-Bob system (Left) and the Alice-antiBob system (Right) as a function of Bob’s position $R_0 = r_0/r_h$ and the quantum parameter $\tilde{\zeta}=\zeta/M$ for different values of $\tilde{\omega}=\omega_i M$.
}
\label{fig7}
\end{figure}
Figs. \ref{fig6} and \ref{fig7} reveal a number of distinctive features of the Dirac field that differ substantially from the scalar case. 
In each figure, the left panels correspond to the Alice-Bob bipartition, while the right panels correspond to the Alice-antiBob bipartition. 
The top and bottom rows display the results for two representative mode frequencies, $\tilde{\omega}=0.1$ and $\tilde{\omega}=0.2$, respectively.

As shown in Figs. \ref{fig6}(a) and \ref{fig6}(c), the logarithmic negativity in the AB bipartition decreases toward a finite limit $\mathcal{N}\simeq0.58$ for any value of the quantum parameter $\tilde{\zeta}$ and mode frequency $\tilde{\omega}$.
This demonstrates that a nonvanishing amount of entanglement survives even when Bob approaches the horizon asymptotically—a well-known characteristic behavior of fermionic fields in both Rindler and Schwarzschild spacetimes \cite{Alsing:2006cj}.
Meanwhile, the entanglement that is lost in the AB bipartition is transferred to the complementary $\text{A}\bar{\text{B}}$ bipartition.
When Bob crosses the horizon, Bob and antiBob coincide, and the entanglement of $\rho_{\text{A}\bar{\text{B}}}$ reaches its maximum with $N(\rho_{\text{A}\bar{\text{B}}})\simeq0.58$, fully compatible with the fermionic entanglement structure.
Turning to mutual information, Figs. \ref{fig7} shows that its qualitative behavior closely follows that of the logarithmic negativity in both bipartitions, and the effects of the quantum corrections remain consistent.
However, in contrast to the logarithmic negativity, the total mutual information obeys a conservation law that was first established in Rindler spacetime.
Importantly, this conservation law continues to hold in the present effective quantum black hole geometry and remains unaffected by the quantum corrections encoded in $\tilde{\zeta}$.
Specifically, for any radial distance $R_0$ and any value of the quantum parameter $\tilde{\zeta}$, one finds that $I_{\text{AB}} + I_{\text{A}\bar{\text{B}}}=2$, which confirms that the redistribution of correlations between the two bipartitions is not modified by the underlying quantum geometry.
Thus, while the logarithmic negativity reflects the transfer of quantum correlations from AB to A$\bar{\text{B}}$ as Bob approaches the horizon, the mutual-information conservation illustrates that the total amount of (quantum + classical) correlations is preserved throughout the process.

Finally, across all the quantum resource measures considered in the effective quantum black hole spacetime, a common feature emerges. 
Although the influence of the quantum parameter $\tilde{\zeta}$ is rather mild for small values, its impact increases sharply once the parameter exceeds the threshold $\zeta\gtrsim 2M$. 
Beyond this point, even a modest increase in $\tilde{\zeta}$ leads to a sudden and significant change in the behavior of all the quantities studied.

\section{Conclusions}\label{sec5}

In this manuscript, we investigate the degradation of quantum entanglement in the third-type black hole solution arising from effective quantum gravity. 
This geometry, which notably possesses no Cauchy horizon, provides an ideal setting for exploring how quantum effects influence quantum information processes in relativistic backgrounds.
Under the near-horizon approximation, the effective quantum black hole geometry can be cast into a Rindler metric. 
This form not only reveals the existence of three distinct timelike Killing vectors but also clarifies the structure of the vacuum modes and their flat-spacetime analogues. 
The resulting correspondence forms the foundation of the framework developed by E. Martín-Martínez and colleagues for investigating entanglement degradation in uniformly accelerated settings.
In the limit of infinite acceleration, the Rindler approximation faithfully reproduces the effective quantum gravity black hole geometry, in which Bob is effectively placed arbitrarily close to the event horizon.

Our analysis demonstrates that the quantum corrections encoded in the third-type black hole significantly modify the behavior of field correlations near the horizon. 
A unified conclusion emerging from our analysis is that the presence of the quantum parameter mitigates the gravity induced degradation of entanglement.
For scalar fields, both the logarithmic negativity and the mutual information exhibit strong sensitivity to the quantum parameter $\tilde{\zeta}$, with even moderate values producing substantial departures from the classical gravitational prediction. 
For Dirac fields, the redistribution of correlations between the AB and $\text{A}\bar{\text{B}}$ bipartitions persists, and the mutual-information conservation law remains intact, indicating that this fermionic kinematic structure is robust against quantum geometric modifications. 
Overall, these results highlight the utility of the third-type effective quantum black hole as a controlled setting for investigating how quantum gravity effects imprint on relativistic quantum information, and they suggest several avenues for further exploration, including dynamical scenarios, higher-spin fields, and extensions to rotating backgrounds.

\acknowledgments
This work is supported by the National Natural Science Foundation of China (12575056), the Young Elite Scientist Sponsorship Program by Guizhou Science and Technology Association (Grant No. GASTYESS202424), the Discipline-Team of Liupanshui Normal University of China (Grant No. LPSSY2023XKTD11), and the Special Fund for Basic Scientific Research of Provincial Universities in Liaoning under grant NO.LS2024Q002, the Fundamental Research Funds for the Central Universities (Grant No. lzujbky-2025-jdzx07), the Natural Science Foundation of Gansu Province (No. 22JR5RA389, No.25JRRA799), and the `111 Center' under Grant No. B20063.

\appendix

%

\end{document}